\documentclass[prd, amsfonts, onecolumn, nofootinbib, showpacs]{revtex4}
\usepackage{graphicx, epsfig}
\usepackage{color}
\usepackage{amsmath}
\usepackage{amssymb}
\newcommand{\be}{\begin{equation}}
\newcommand{\ee}{\end{equation}}
\newcommand{\bea}{\begin{eqnarray}}
\newcommand{\eea}{\end{eqnarray}}

\newcommand{\gapp}{\mathrel{\raise.3ex\hbox{$>$}\mkern-14mu
\lower0.6ex\hbox{$\sim$}}}
\newcommand{\lapp}{\mathrel{\raise.3ex\hbox{$<$}\mkern-14mu
\lower0.6ex\hbox{$\sim$}}}
\def\bbox{{\,\lower0.9pt\vbox{\hrule \hbox{\vrule height 0.2 cm
\hskip 0.2 cm \vrule  height 0.2 cm}\hrule}\,}}

\begin{document}
\title{Multibrane DGP model: Our universe as a stack of $(2+1)$-dim branes}
\author{}
\affiliation{}
\author{De-Chang Dai$^{1,}{}^3$, Dejan Stojkovic$^2$, Bin Wang$^3$ and Cheng-Yong Zhang$^3$}

\address{$^1$Institute of Natural Sciences,
Shanghai Jiao Tong University, Shanghai 200240, China }

\address{$^2$HEPCOS, Department of Physics, SUNY at Buffalo, Buffalo, NY 14260-1500\\
and  Perimeter Institute for Theoretical Physics, 31 Caroline St. N., Waterloo, ON, N2L 2Y5, Canada }

\address{$^{3}$Center for Astrophysics and Astronomy, Department of Physics and Astronomy,
Shanghai Jiao Tong University, Shanghai 200240, China }


\begin{abstract}
\widetext
We consider scenario in which our $(3+1)$-dim universe is actually a dense stack of multiple parallel $(2+1)$-dim branes. For this purpose, we generalize the DGP model to a multi-brane case. We solve for the propagation of the scalar field and gravity in this setup. At short distances (high momenta) along the branes interactions follow  the $(2+1)$-dim laws, while at large distances (low momenta) interactions follow the usual $(3+1)$-dim laws. This feature is inherited from the original DGP model. In the direction perpendicular to the brane, we show that interactions become $(3+1)$-dim at low momenta which are unable to resolve the interbrane separation. Thus,  this is one of the explicit  constructs of the ``vanishing dimensions" scenario where high energy physics appears to be lower dimensional rather than higher dimensional.
\end{abstract}


\maketitle

\section{Introduction}

An idea that our universe might be higher dimensional on large cosmological scales was explored in the so-called Dvali-Gabadadze-Porrati (DGP) model \cite{dgp} (see also cascading gravity in \cite{rvhk,rkt,rhkt,Kaloper:2007ap,Kaloper:2007qh,CuadrosMelgar:2007jx}). Basically, our universe can be represented by a brane embedded in a higher dimensional bulk. The fine tuning between the brane tension and  cosmological constant in the bulk allows gravity to be $(3+1)$-dim on short distances (shorter than some cross-over scale) and $(4+1)$-dim on large distances.

\begin{figure}[h!]
   \centering
\includegraphics[width=8cm]{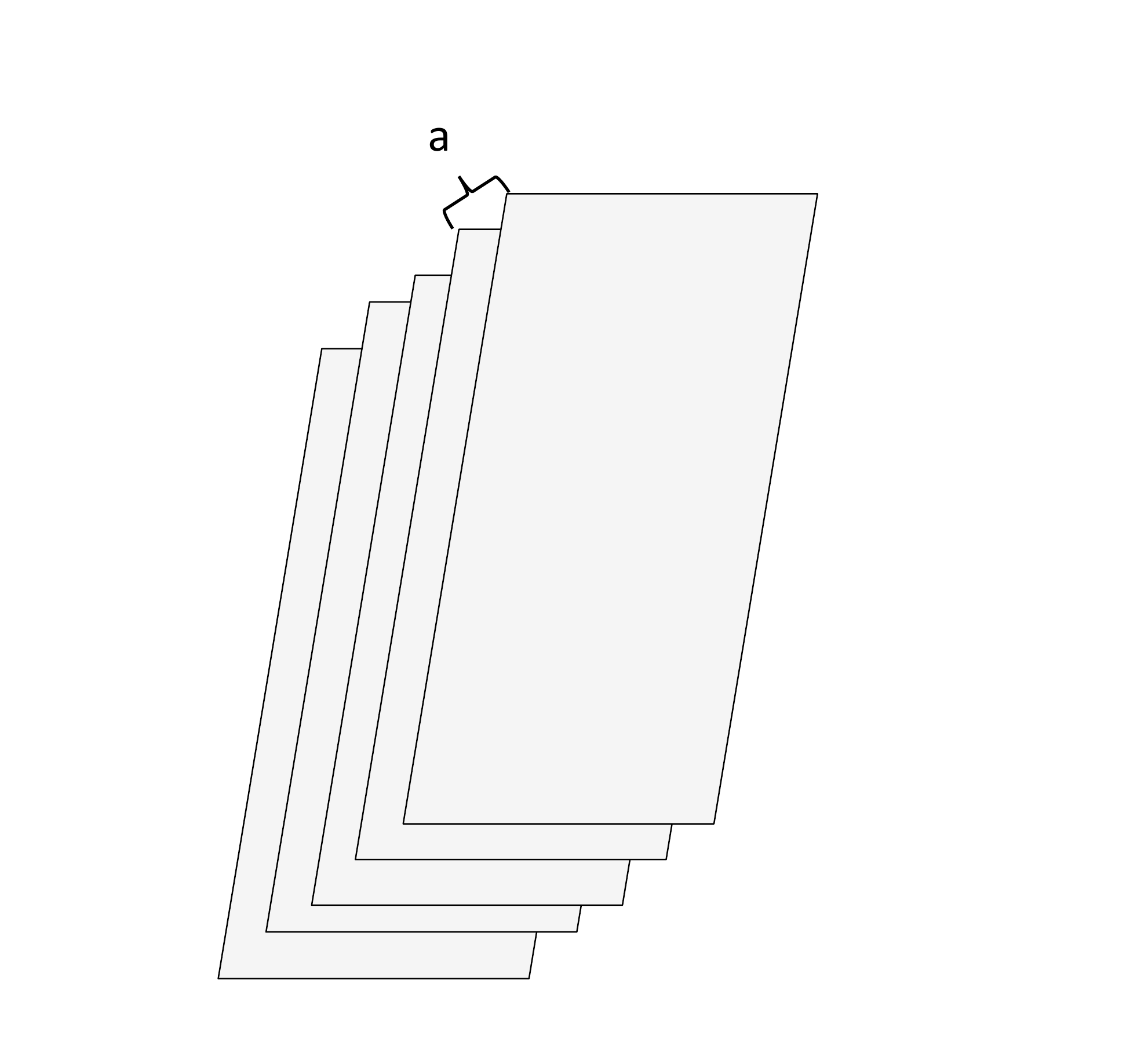}
\caption{The $(3+1)$-dim universe that we live in could be comprised of the stack of densely packed $(2+1)$-dim  branes. If the inter-brane separation corresponds to the dimensional cross-over scale ($a$ or smaller), our $(3+1)$-dim universe would appear smooth at energies smaller than $1/a$.}
\label{stack}
\end{figure}

On the other hand, there is a strong motivation to reduce the number of dimensions to less than $3+1$ on ultra-short scales \cite{Anchordoqui:2010er,Anchordoqui:2010hi,Stojkovic:2013lga,Stojkovic:2014lha,Mureika:2011bv,as}. An explicit extension of the DGP model to lower dimensions  was constructed in \cite{Hao:2014tsa}. However, obtaining a realistic model of our space-time which is lower dimensional at short scales is more involving than a simple extension of the DGP model. In particular it is not enough to have only one $(2+1)$-dim brane embedded in a $(3+1)$-dim bulk. In order to have a smooth $(3+1)$-dim universe in the infrared (IR), one needs a setup with a stack of $(2+1)$-dim branes, e.g. as shown in Fig.~\ref{stack}.
In this paper we concentrate on this scenario. We are trying to construct a $(3+1)$-dim universe which on short scales appears to be $(2+1)$-dim, while on large distances we recover a usual $(3+1)$-dim behavior (in all directions, and not only along one direction like in simple generalizations of the DGP models).

\section{Free scalar field on the stack of branes}
\label{fsf}

We first consider a free scalar field $\phi$ on the non-trivial background represented by the stack of parallel branes as shown in Fig.~\ref{stack}.
The bulk space-time is $(3+1)$-dim with Cartesian coordinates $(t,x,y,z)$. In this bulk, we have a stack of $(2+1)$-dim branes with coordinates $(t,x,y)$, which are located at the points labeled by $z=na$. Here, $n$ is an integer, while $a$ is the distance between branes.  The scalar field  Lagrangian in this background is
\begin{equation}
\mathcal{L}=-\frac{M_4^2}{2}\partial_\mu \phi \partial^\mu \phi -\sum_{n=-\infty}^\infty \frac{M_3}{2}\delta (z-na)\partial_i \phi \partial^i \phi ,
\end{equation}
where the Latin index $i$ goes over the coordinates in the $(2+1)$-dim brane $(t,x,y)$, while the Greek index $\mu$ goes over the coordinates in the whole $(3+1)$-dim space $(t,x,y,z)$. $M_3$ is the tension on $(2+1)$-dim branes (they all have the same tension), while $M_4$ is the bulk cosmological constant in the full $(3+1)$-dim space. The scalar field equation of motion is
\begin{equation}
\label{eom}
m_3\partial_\mu \partial^\mu \phi +\sum_{n=-\infty}^\infty \delta (z-na)\partial_i \partial^i \phi =0 ,
\end{equation}
where $m_3=M_4^2/M_3$. Consider now a free particle solution in the following form
\begin{equation}
\phi=A\exp(-iEt+iP_x x+iP_y y)u(z)
\end{equation}
After substituting this ansatz into  Eq.~(\ref{eom}), one finds
\begin{equation}
-k^2u -\partial_z^2 u -\sum_{n=-\infty}^\infty\frac{k^2}{m_3}\delta(z-na)u=0.
\end{equation}
Here, $k^2=E^2-P_x^2-P_y^2$. This is the well known Dirac-Kronig-Penny model. According to Bloch's theorem, if the solution has translational symmetry, then it can be represented with a Bloch wave
\begin{equation}
u=\exp(iQz)\nu_Q (z) .
\end{equation}
The non-trivial solution is chosen in the first Brillouin zone, $-\pi/a<Q<\pi/a$, while $\nu$ is a periodic function in the $z$-direction, $\nu_Q(z+a)=\nu_Q(z)$.
Thus, a free scalar field can propagate in the bulk like a traveling wave except that the amplitude is given by a periodic function.
In the next step, we will find the solution which is sourced by a unit source on the brane.

\section{Static scalar field sourced by the matter on brane}
\label{ssf}

In order to find a general solution for the scalar field generated by some source, we first have to find the solution sourced by a unit source. This solution will give us the basic properties like effective dimensionality of space-time.  Then, if we want, by integration we can find a solution for an arbitrary source. Consider a unit matter source located at $(x=0,y=0,z=0)$. This source will generate a scalar field according to
\begin{equation}
\label{eom1}
M_4^2\partial_\mu \partial^\mu \phi +M_3\sum_{n=-\infty}^\infty \delta (z-na)\partial_i \partial^i \phi =\delta(x)\delta(y)\delta(z)
\end{equation}
We can write down the solution for $\phi$ in the form
\begin{equation}\label{phi}
\phi=\int \frac{1}{(2\pi)^2}dk_x dk_y G(k_x,k_y,z)\exp(ik_x x+ik_y y)
\end{equation}
where $G(k_x,k_y,z)$ is an unknown function that needs to be determined from the equation of motion.
Eq.~(\ref{eom1}) becomes
\begin{equation}
\label{green1}
M_4^2(k^2-\partial_z^2) G +M_3\sum_{n=-\infty}^\infty \delta (z-na)k^2 G =\delta(z)
\end{equation}
where $k^2=k_x^2+k_y^2$. Throughout this paper, $k$ will always represent momentum along the brane. If the scalar field is not on the brane, the equation is simply
\begin{equation}
(k^2-\partial_z^2) G=0
\end{equation}
In this case, the solution for $G$ could be written as the combination of $\exp(kz)$ and $\exp(-kz)$. Therefore, we can write an equation representing $G$ as
\begin{eqnarray}
&&G=A_j \exp(-k(z-ja))+B_j \exp(k(z-ja))\\
&&ja<z<(j+1)a
\end{eqnarray}
Here, $j$ is an integer. On the $z=0$ brane, the solution must satisfy
\begin{equation}
\label{0brane}
2M_4^2 (A_0-B_0)k +M_3 k^2 (A_0+B_0)=1
\end{equation}
Here we imposed the discrete $z_2$ symmetry on $z=0$ brane. For other branes, located at $z\neq 0$, the following boundary conditions must be satisfied
\begin{eqnarray}
&& A_j \exp(-ka)+B_j \exp(ka)=A_{j+1}+B_{j+1}\\
&&M_4^2\Big(A_{j+1}-B_{j+1})-(A_j\exp(-ka)-B_j\exp(ka))\Big)+M_3 k (A_j \exp(-ka)+B_j \exp(ka))=0 .
\end{eqnarray}
These equations can be written as
\begin{eqnarray}
&&\left[ {\begin{array}{c}
   A_{j+1} \\
   B_{j+1}\\
  \end{array} } \right]
=M
\left[ {\begin{array}{c}
   A_{j} \\
   B_{j}\\
  \end{array} } \right]\\
&&M=  \left[ {\begin{array}{cc}
   (1-r_0k)\exp(-ka) & -r_0 k\exp(ka) \\
   r_0 k\exp(-ka) & (1+r_0k)\exp(ka) \\
  \end{array} } \right]
\end{eqnarray}
where $r_0 =\frac{M_3}{2M_4^2}$. In fact, $r_0$ represents the dimensional crossover scale along the brane, i.e. on distances shorter than $r_0$ the field propagates in a $(2+1)$-dim spacetime, while on distances larger than $r_0$ one recovers $(3+1)$-dim behavior.  We can also rewrite these equations as
\begin{eqnarray}
&&\left[ {\begin{array}{c}
   A_{j} \\
   B_{j}\\
  \end{array} } \right]
=M^j
\left[ {\begin{array}{c}
   A_{0} \\
   B_{0}\\
  \end{array} } \right]\\
\end{eqnarray}
The eigenvalues of the matrix $M$ are
\begin{eqnarray}
\lambda_1&=&\frac{F-\sqrt{F^2-4}}{2}<1\\
\lambda_2&=&\frac{F+\sqrt{F^2-4}}{2}>1 \\
{\rm with \ \ \ }F&=&(1-r_0k)\exp(-ka)+(1+r_0k)\exp(ka)
\end{eqnarray}
Since $\lambda_2>1$, if
$\left[ {\begin{array}{c}
   A_{0} \\
   B_{0}\\
  \end{array} } \right]$ includes any component of the eigenvector that corresponds to $\lambda_2$, it will become divergent as  $j\rightarrow \infty$. Therefore only  eigenvector corresponding to  $\lambda_1$ can be included in the matrix
  $\left[ {\begin{array}{c}
   A_{0} \\
   B_{0}\\
  \end{array} } \right]$.
  Therefore, we can write

\begin{equation}
\label{solu0}
\left[ {\begin{array}{c}
   A_{0} \\
   B_{0}\\
  \end{array} } \right]= H \left[ {\begin{array}{c}
   r_0 k\exp(ka) \\
   (1-r_0k)\exp(-ka)-\lambda_1\\
  \end{array} } \right]
\end{equation}
One can normalize the above matrix equation with Eq.~(\ref{0brane}) and find the normalization constant
\begin{equation}
H=\frac{1}{2M_4^2 k }\frac{1}{(r_0k+r_0^2k^2)\exp(ka)-(1-r_0k)^2\exp(-ka)+(1-r_0k)\lambda_1}
\end{equation}
We can now find the approximate solutions for the field $\phi$ on different branes
\begin{itemize}

\item On the $z=0$ brane:

Since parameter $a$ is the inter-brane distance, it is instructive to see what happens at much larger and much smaller scales.
In the limit of $k>>1/a$, i.e. large momenta capable of probing the inter-brane distance so that individual branes are resolved, one finds

\begin{eqnarray}
H&\simeq& \frac{1}{2M_4^2 k }\frac{1}{(r_0k+r_0^2k^2)\exp(ka)} \\
\left[ {\begin{array}{c}
   A_{0} \\
   B_{0}\\
  \end{array} } \right]&\simeq& H \left[ {\begin{array}{c}
   r_0 k\exp(ka) \\
   0\\
  \end{array} } \right] \\
G&\simeq&\frac{1}{2M_4^2 }\frac{1}{(k+r_0k^2)} .
\end{eqnarray}
This is the same as in the single brane case \cite{Hao:2014tsa}. We further have the parameter $r_0$ as the dimensional crossover scale along the brane. If $r<<r_0$ or $k>>1/r_0$, then $G\simeq \frac{1}{2M_4^2 }\frac{1}{r_0k^2}$. From Eq.~(\ref{phi}) we find $\phi\sim \ln(r)$, which is the expected $(2+1)$-dim behavior. Thus, at short distances or high momenta along the brane, a scalar particle effectively propagates in a $(2+1)$-dim space.  If $r>>r_0$ or $k<<1/r_0$, then $G\simeq \frac{1}{2M_4^2 }\frac{1}{k}$, which gives $\phi\sim 1/r$. This is the usual $(3+1)$-dim behavior.
So, in the IR, at large distances or low momenta along the brane, a scalar particle effectively propagates in the full $(3+1)$-dim space.

If $k<<1/a$ and $k<<1/r_0$ (low momenta which can't resolve individual branes nor the dimensional crossover along the brane), one finds
\begin{eqnarray}
F&\simeq&  2+(a^2+2r_0a)k^2\\
\lambda_1&\simeq&  1-\sqrt{a^2+2r_0a}k\\
H&\simeq& \frac{1}{2M_4^2 k }\frac{1}{(2r_0+a-\sqrt{a^2+2r_0a})k} \\
\left[ {\begin{array}{c}
   A_{0} \\
   B_{0}\\
  \end{array} } \right]&\simeq& H \left[ {\begin{array}{c}
   r_0 k \\
  (-r_0-a+\sqrt{a^2+2r_0a})k\\
  \end{array} } \right] \\
G&=&\frac{1}{2M_4^2 k}\frac{a}{\sqrt{a^2+2ar_0}}
\end{eqnarray}
 As expected, $\phi\sim 1/r$ in this case.

\item On the $z=ja$ brane and $(x=0,y=0)$:

When $k>>1/a$, then $\lambda_1\sim \frac{1}{k}\exp(-ka)$. Therefore, the eigenvalue $\lambda_1$ is highly suppressed for large $k$. This means that we do not need to integrate the large $k$ modes, because their contribution disappears very quickly. The only modes surviving in the $dk_xdk_y$ integration will be the low $k$ modes.  We then focus on $k<<1/a$. If we also require $k<<1/r_0$, we get
\begin{eqnarray}
\lambda_1&\simeq&  1-\sqrt{a^2+2r_0a}k\\
G&\simeq&\frac{1}{2M_4^2 k}\frac{a}{\sqrt{a^2+2ar_0}}\lambda_1^j\simeq \frac{1}{2M_4^2 k}\frac{a}{\sqrt{a^2+2ar_0}} \exp(-z\sqrt{1+2r_0/a}k)
\end{eqnarray}
Here, we replaced $ja$ with $z$ and used an approximation $(1-x)^j\simeq \exp(-jx)$. After integrating over $dk_xdk_y$, we get $\phi\sim 1/z$ in the case. Thus, in the direction perpendicular to the brane, we recover a usual $3+1$-dim behavior at low momenta.

\end{itemize}

\section{Gravity on the stack of branes }

In previous sections we derived the behaviour of the free and sourced scalar field on the stack of branes and saw that it has the desired properties.
In this section we take gravity into consideration.
The gravitational Lagrangian in our setup with a stack of multiple parallel branes is
\begin{equation}
\label{Lg}
\mathcal{L}=\frac{M_4^2}{2}R_4+\sum_j \delta(z-ja) \frac{M_3}{2} R^j + \sum_j \delta(z-ja) L^j .
\end{equation}
Here, $R_4$ is the $(3+1)$-dim scalar curvature, $R^j$ is the $(2+1)$-dim scalar curvature on the $z=ja$ brane, while $L^j$ is the matter Lagrangian on the $z=ja$ brane. Matter is constrained on the branes, and it does not have momentum in $z$-direction.

Gravitational excitations around the flat space are described by
\begin{equation}
g_{\mu\nu}=\eta_{\mu\nu}+h_{\mu\nu}
\end{equation}
where $\eta_{\mu\nu}$ is the Minkowski metric, while  $h_{\mu\nu}$ are small perturbations. Greek letters go over all the indices $(t,x,y,z)$.

We first consider matter distribution which is localized at the $z=0$ brane, and calculate induced gravity it produces. In this case, the energy-momentum tensor can be written as
\begin{equation}
T_{\mu\nu}=\left[ {\begin{array}{cc}
   \delta(z)T^0_{kl} & 0\\
   0 & 0\\
  \end{array} } \right]
\end{equation}
 where the lower case Latin indices  go over $(t,x,y)$. We impose the harmonic gauge $\partial^\nu h_{\mu\nu} =\frac{1}{2}\partial_\mu h^\nu_\nu$ in the $(3+1)$-dim spacetime. The equations of motion take the form
\begin{equation}
\label{eqg}
\frac{M_4^2}{2}G_{\mu\nu}+\sum_j\frac{\delta(z-ja)}{2}M_3G^3_{lk}\delta^l_\mu\delta^k_\nu=\frac{\delta(z)}{2}T^0_{lk}\delta^l_\mu\delta^k_\nu .
\end{equation}
where $G_{\mu\nu}$ is the $(3+1)$-dim Einstein tensor, while  $G^3_{lk}$ is the $(2+1)$-dim Einstein tensor. In the linearized form, these two quantities can be written as
\begin{eqnarray}
G_{\mu\nu}&=&-\frac{1}{2}(\Box_4 h_{\mu\nu}-\frac{1}{2}\eta_{\mu\nu} \Box h)\\
G^3_{lk}&=&-\frac{1}{2}(\Box_3 h_{lk}-\frac{1}{2}\eta_{lk} \Box h^3)\\
h&=&\eta^{\mu\nu}h_{\mu\nu}\\
h^3&=&\eta^{ij}h_{ij}
\end{eqnarray}
In \cite{Hao:2014tsa}, it was shown that if $T_{\mu z}=0$, then $h_{\mu z}=0$ is consistent with the gauge condition in the single brane case. This condition is valid in the multibrane case too. Therefore we will set $h_{\mu z}=0$ in our equations. Then Eq.~(\ref{eqg}) simplifies to
\begin{equation}
\label{eqg2}
-\frac{1}{2}(M_4^2\Box+M_3\sum_j\delta(z-ja)\Box_3)h_{lk}=(T_{lk}-\frac{1}{2}\eta_{lk}T)\delta(z)-M_3\sum_j \partial_l\partial_k \delta(z-ja)h^4_4 .
\end{equation}
We can now perform the Fourier transform on $h_{\mu\nu}$ and $T_{\mu\nu}$ as
\begin{eqnarray}
h_{lk}&=&\int \frac{dp^3}{2\pi^3}\exp(ip_tt+ip_xx+ip_yy) \hat{h}_{lk} (\vec{P},z)\\
T_{lk}&=&\int \frac{dp^3}{2\pi^3}\exp(ip_tt+ip_xx+ip_yy) \hat{T}_{lk}(\vec{P},z)\\
T'_{lk}&=&\int \frac{dp^3}{2\pi^3}\exp(ip_tt+ip_xx+ip_yy) \hat{T'}_{lk}(\vec{P},z)
\end{eqnarray}
where $\vec{P}=(p^t,p^x,p^y)$. Here, another energy-momentum tensor $T'_{\mu\nu}$ is introduced since we want to calculate the gravitational potential of $T'_{\mu\nu}$ generated by the source $T_{\mu\nu}$ (in other words, we want to calculate the gravitational force between two mass distributions). Since in our case both $T'_{\mu\nu}$ and $T_{\mu\nu}$ are localized on the $z=ja$ brane, their energy-momentum conservation to the linear leading order is written as
\begin{equation}
p_l\hat{T}^{lk}=p_k\hat{T}^{lk}=p_l\hat{T'}^{lk}=p_l\hat{T'}^{lk}=0 .
\end{equation}
After multiplying  Eq.~(\ref{eqg2}) with $\hat{T'}^{lk}$, one finds that
\begin{equation}
\frac{1}{2}\Big(M_4^2(P^2-\partial_z^2)+M_3 \sum_j\delta(z-ja)P^2\Big)\hat{h}_{lk}\hat{T'}^{lk}=-(\hat{T}_{lk}\hat{T'}^{lk}-\frac{1}{2}\hat{T}^l_l\hat{T'}^k_k)\delta(z)
\end{equation}
where $P^2=-p_lp^l$.
This equation is exactly the same as Eq.~(\ref{green1}), except for the multiplicative constant $-2(\hat{T}_{lk}\hat{T'}^{lk}-\frac{1}{2}\hat{T}^l_l\hat{T'}^k_k)$. Therefore, we can find
\begin{eqnarray}
&&\hat{h}_{lk}\hat{T'}^{lk}=-2(\hat{T}_{lk}\hat{T'}^{lk}-\frac{1}{2}\hat{T}^l_l\hat{T'}^k_k)\times( A_0 \exp(-Pz)+B_0\exp(Pz))\\
&&0\le z<a
\end{eqnarray}
Constants $A_0$ and $B_0$ could be found from Eq.~(\ref{solu0}). If for simplicity we take $a>>r_0$, after performing an inverse Fourier transform, one finds that the gravitational potential
\begin{equation}
V(r) \sim 1/r {\rm \ \ \ as \ \ } r>>r_0 {\rm \ \ \ or \ \ }  k<<1/r_0,
\end{equation}
and
\begin{equation}
V(r)\sim \ln(r) {\rm \ \ \ as \ \ }   r<<r_0 .
\end{equation}
The last two equations indicate that at large distances or low momenta (along the brane), gravity has the $(3+1)$-dim behavior, while at short distances it has the $(2+1)$-dim behavior.

In the z-direction (perpendicular to the brane), large momentum modes $k>>1/a$ are highly suppressed by a factor $exp(-jka)$. Therefore, the large $k$ modes decay very quickly in the $z$-direction. Therefore, we consider only $k<<1/a$ modes, for which we get
\begin{equation}
V(z) \sim 1/z {\rm \ \ \ as \ \ }   k<<1/r_0,
\end{equation}
 We emphasis that this result is valid only if matter is localized on the branes, i.e. both matter distributions $T'$ and $T$ do not have any component in the $z$-direction. If matter is allowed to move in $z$-direction, then $h_{\mu z}$ is not zero and this result does not apply. However in this case the $(3+1)$-dim universe is no longer described by a stack of $(2+1)$-dim branes with localized matter.

\section{Conclusion}
Motivation for the work presented here is twofold. The first one is an extension of the DGP model to a multibrane case. If we live in a universe (or a $(3+1)$-dim brane) which is itself comprised of a dense stack of parallel branes, then we have two scales: the DGP dimensional crossover scale $r_0$ and the interbrane separation $a$. The original DGP setup implies that the interactions will be lower dimensional for high and $(3+1)$-dim  for low momenta (i.e. short and large distances), at least along the brane(s). We verified that this is true by explicitly solving the cases of the scalar field and gravity. In addition, we showed that this feature remains true even in the direction perpendicular to the brane(s).  At low momenta, when the wavelength of the force mediator is larger than the inter-brane distance and unable to resolve the individual branes, interactions appear to be $(3+1)$-dim. These properties fit very well into the recently proposed ``vanishing dimensions" scenario where physics at short scales appears to be lower dimensional, which was another explicit motivation for this work.

 In section \ref{fsf} we showed that a free scalar field can propagate in the bulk like a traveling wave except that the amplitude is given by a periodic function.
 In contrast, in section \ref{ssf} we considered a static scalar field sourced by the matter on brane and showed that it has the desired dimensional properties as described above.

A caveat that is left in this setup is that for the case of gravitational interactions we assumed that the source (gravitating matter) was localized on the branes and it did not have any momentum in the perpendicular dimension (calculations significantly simplify in this case). We obtained the desired behavior of the interactions, but in a realistic setup, matter fields in IR should be able to propagate in all directions, not only along the branes. We can remedy this situation by allowing matter to move off the brane classically by replacing the terms of the type  $\delta(z) L_{\rm matter} $ in the Lagrangian (\ref{Lg}) with $f(t,x,y,z)+\delta(z) g(t,x,y,z)$ which would allow matter to leak into the bulk.
 However, from the experience with the Randall-Sundrum models \cite{Dubovsky:2000am,Gregory:2000rh,Rubakov:2001kp}, we know that matter would leak to the bulk anyway. The reason is that the wave function for the localized matter does not vanish exactly off the brane. If the next parallel brane is close, then matter could easily propagate from a brane to a brane.

\begin{acknowledgments}
This work was partially supported by the US National Science Foundation, under Grant No. PHY-1066278 and PHY-1417317, by the Science and Technology Commission of Shanghai Municipality under grand No. 14ZR1423200 and 11DZ2260700, and by NNSFC.
\end{acknowledgments}


\begin{thebibliography}{99}



\bibitem{dgp}
G.Dvali, G. Gabadadze, M. Porrati, \emph{Phys. Lett. B} {\bf 485} (2000) 208-214;
  R.~Gregory, V.~A.~Rubakov and S.~M.~Sibiryakov,
  Phys.\ Rev.\ Lett.\  {\bf 84}, 5928 (2000)
  [hep-th/0002072].



\bibitem{rvhk}
C. de Rham, G. Vali, S. Hofmann, J. Khoury, \emph{Phys. Rev. Lett.} {\bf 100}, 251603 (2008), arxiv: hep-th/0711.2072


\bibitem{rkt}
C. de Rham, J. Khoury, A. Tolley, \emph{Phys. Rev. Lett.} {\bf 103}, 161601 (2009), arxiv: hep-th/0907.0473v2

\bibitem{rhkt}
C. de Rham, S. Hofmann, J. Khoury, A. Tolley, \emph{JCAP} {\bf 0802:011}, (2008), arxiv: hep-th/0712.2821

\bibitem{Kaloper:2007ap}
  N.~Kaloper and D.~Kiley,
  JHEP {\bf 0705}, 045 (2007)
  [hep-th/0703190];

\bibitem{Kaloper:2007qh}
  N.~Kaloper,
  Mod.\ Phys.\ Lett.\ A {\bf 23}, 781 (2008)
  [arXiv:0711.3210 [hep-th]].


\bibitem{CuadrosMelgar:2007jx} 
  B.~Cuadros-Melgar, E.~Papantonopoulos, M.~Tsoukalas and V.~Zamarias,
  Phys.\ Rev.\ Lett.\  {\bf 100}, 221601 (2008)
  [arXiv:0712.3232 [hep-th]].

\bibitem{Anchordoqui:2010er}
  L.~Anchordoqui, D.~C.~Dai, M.~Fairbairn, G.~Landsberg and D.~Stojkovic,
  Mod.\ Phys.\ Lett.\ A {\bf 27}, 1250021 (2012)
  [arXiv:1003.5914 [hep-ph]].





\bibitem{Anchordoqui:2010hi}
  L.~A.~Anchordoqui, D.~C.~Dai, H.~Goldberg, G.~Landsberg, G.~Shaughnessy, D.~Stojkovic and T.~J.~Weiler,
  Phys.\ Rev.\ D {\bf 83}, 114046 (2011)
  [arXiv:1012.1870 [hep-ph]].

\bibitem{Stojkovic:2013lga}
  D.~Stojkovic,
  arXiv:1304.6444 [hep-th].


\bibitem{Stojkovic:2014lha}
  D.~Stojkovic,
  Mod.\ Phys.\ Lett.\ A {\bf 28}, 1330034 (2013)
  [arXiv:1406.2696 [gr-qc]].



\bibitem{Mureika:2011bv}
  J.~R.~Mureika and D.~Stojkovic,
  Phys.\ Rev.\ Lett.\  {\bf 106}, 101101 (2011)
  [arXiv:1102.3434 [gr-qc]];
  Phys.\ Rev.\ Lett.\  {\bf 107}, 169002 (2011)
  [arXiv:1109.3506 [gr-qc]].

\bibitem{as}
  N.~Afshordi and D.~Stojkovic,
  arXiv:1405.3297 [hep-th].

\bibitem{Hao:2014tsa}
  P.~Hao and D.~Stojkovic,
  arXiv:1404.7145 [gr-qc].

\bibitem{Dubovsky:2000am}
  S.~L.~Dubovsky, V.~A.~Rubakov and P.~G.~Tinyakov,
  Phys.\ Rev.\ D {\bf 62}, 105011 (2000)
  [hep-th/0006046].

\bibitem{Gregory:2000rh}
  R.~Gregory, V.~A.~Rubakov and S.~M.~Sibiryakov,
  Class.\ Quant.\ Grav.\  {\bf 17}, 4437 (2000)
  [hep-th/0003109].

\bibitem{Rubakov:2001kp}
  V.~A.~Rubakov,
  Phys.\ Usp.\  {\bf 44}, 871 (2001)
  [Usp.\ Fiz.\ Nauk {\bf 171}, 913 (2001)]
  [hep-ph/0104152].



\end{thebibliography}
\end{document}